\documentclass[twocolumn,english,aps,prb,twocolumn,superscriptaddress,bibnotes,amsmath,amssymb,floatfix]{revtex4-1}
\usepackage[colorlinks=true,citecolor=blue,linkcolor=magenta]{hyperref}

\usepackage[utf8]{inputenc}
\usepackage[english]{babel}
\usepackage{amsmath,amsfonts,amssymb}
\usepackage[T1]{fontenc}
\usepackage{url}
\usepackage{soul}
\usepackage{changes}

\usepackage{amsmath}
\usepackage{amsfonts}
\usepackage{amssymb}
\usepackage[version=3]{mhchem}
\usepackage{stmaryrd}
\usepackage{epstopdf}
\usepackage{graphicx}
\graphicspath{{./Figures/}}

\begin{document}

\title{Reconfigurable radiofrequency filters based on versatile soliton microcombs}

\author{Jianqi Hu}
\thanks{These authors contributed equally to the work}
\affiliation{Institute of Physics, Swiss Federal Institute of Technology Lausanne (EPFL), Photonic Systems Laboratory (PHOSL), STI-IEL, Lausanne CH-1015, Switzerland.}

\author{Jijun He}
\thanks{These authors contributed equally to the work}
\affiliation{Institute of Physics, Swiss Federal Institute of Technology Lausanne (EPFL), Laboratory of Photonics and Quantum Measurements (LPQM), SB-IPHYS, Lausanne CH-1015, Switzerland.}

\author{Junqiu Liu}
\affiliation{Institute of Physics, Swiss Federal Institute of Technology Lausanne (EPFL), Laboratory of Photonics and Quantum Measurements (LPQM), SB-IPHYS, Lausanne CH-1015, Switzerland.}

\author{Arslan S. Raja}
\affiliation{Institute of Physics, Swiss Federal Institute of Technology Lausanne (EPFL), Laboratory of Photonics and Quantum Measurements (LPQM), SB-IPHYS, Lausanne CH-1015, Switzerland.}

\author{Maxim Karpov}
\affiliation{Institute of Physics, Swiss Federal Institute of Technology Lausanne (EPFL), Laboratory of Photonics and Quantum Measurements (LPQM), SB-IPHYS, Lausanne CH-1015, Switzerland.}

\author{Anton Lukashchuk}
\affiliation{Institute of Physics, Swiss Federal Institute of Technology Lausanne (EPFL), Laboratory of Photonics and Quantum Measurements (LPQM), SB-IPHYS, Lausanne CH-1015, Switzerland.}

\author{Tobias J. Kippenberg}
\email[]{tobias.kippenberg@epfl.ch}
\affiliation{Institute of Physics, Swiss Federal Institute of Technology Lausanne (EPFL), Laboratory of Photonics and Quantum Measurements (LPQM), SB-IPHYS, Lausanne CH-1015, Switzerland.}

\author{Camille-Sophie Br\`es}
\email[]{camille.bres@epfl.ch}
\affiliation{Institute of Physics, Swiss Federal Institute of Technology Lausanne (EPFL), Photonic Systems Laboratory (PHOSL), STI-IEL, Lausanne CH-1015, Switzerland.}

\maketitle

\noindent\textbf{\noindent
The rapidly maturing integrated Kerr microcombs show significant potential for microwave photonics. Yet, state-of-the-art microcomb based radiofrequency (RF) filters have required programmable pulse shapers \cite{xue2014programmable,xu2019advanced,xu2019high}, which inevitably increase the system cost, footprint, and complexity. Here, by leveraging the smooth spectral envelope of single solitons, we demonstrate for the first time microcomb based RF filters free from any additional pulse shaping. More importantly, we achieve all-optical reconfiguration of the RF filters by exploiting the intrinsically rich soliton configurations. Specifically, we harness the perfect soliton crystals \cite{karpov2019dynamics} to multiply the comb spacing thereby dividing the filter passband frequencies. Also, a completely novel approach based on the versatile interference patterns of two solitons within one round-trip  \cite{wang2017universal}, enables wide reconfigurability of RF passband frequencies according to their relative azimuthal angles. The proposed schemes demand neither an interferometric setup nor another pulse shaper for filter reconfiguration, providing a practical route towards chipscale, widely reconfigurable microcomb based RF filters.
}

Thanks to the ever-maturing photonic integration, RF photonic systems and subsystems have been brought to new height\cite{capmany2007microwave,marpaung2019integrated}, in terms of footprint, scalability, and potentially cost-effectiveness. Particularly, RF filtering towards chip-scale is a key enabling function \cite{sancho2012integrable,metcalf2016integrated,zhuang2015programmable,marpaung2013si,eggleton2019brillouin,fandino2017monolithic,xue2014programmable,xu2019advanced,xu2019high}. Paradigm demonstrations include the integration of the basic filtering blocks, such as delay lines \cite{sancho2012integrable}, optical spectral shaper \cite{metcalf2016integrated}, programmable mesh topologies \cite{zhuang2015programmable}, ring resonators \cite{marpaung2013si}, as well as the use of stimulated Brillouin scattering (SBS) in waveguides \cite{eggleton2019brillouin}. Recently, an all-integrated RF photonic filter has  been shown in a monolithic platform \cite{fandino2017monolithic}. 
Among others, RF filters constructed from tapped delay line (TDL) structures attract great attention. They can be classified into two types \cite{capmany2006tutorial}, depending on whether the filtering profile is given by the physical path delays \cite{zhuang2015programmable,fandino2017monolithic} or the light source spectra \cite{supradeepa2012comb,sancho2012integrable,metcalf2016integrated,maram2019discretely,xue2014programmable,xu2019advanced,xu2019high}. While the former approach is straightforward, a multi-wavelength source combined with dispersive propagation also functions as a TDL filter. This greatly simplifies the structure complexity of a finite impulse response (FIR) filter, as only a single dispersive delay line is needed. Nevertheless, the main complexity is then shifted to the multi-wavelength source. Electro-optic combs \cite{supradeepa2012comb,metcalf2016integrated,zhu2017novel} or mode-locked lasers \cite{maram2019discretely} are generally adopted as light sources, which remain expensive and bulky options. 

Integrated optical Kerr combs (microcombs) have appeared as an interesting alternative. Indeed, microcombs have already been applied not only to filter RF signals \cite{xue2014programmable,xu2019advanced,xu2019high}, but also for various RF photonic processing, such as true-time delay beamforming \cite{xue2018microcomb}, RF channelization \cite{xu2018broadband}, and analog computation \cite{tan2019microwave}. The large comb spacing of microcombs also enhances RF filters with broader Nyquist zone (spur free range), lower latency \cite{xue2014programmable}, less dispersion induced fading, as well as larger number counts of equivalent delay lines \cite{xu2019advanced}, unparalleled by other approaches. However, so far all these microcomb based RF filters have been implemented on either dark pulses \cite{xue2014programmable,xue2015mode} or complex soliton crystal states \cite{xu2019advanced,xu2019high}. Additional programmable pulse shaping modules are inevitably required to equalize or smooth the comb spectral shape. Thus, the system complexity is significantly increased while the potential for low-cost and high-volume applications is compromised. To date, harnessing the smooth ${\rm sech^2}$ spectral envelope of single soliton \cite{herr2014temporal,stern2018battery,raja2019electrically} or other regulated dissipative Kerr soliton (DKS) states for photonic RF filtering is yet to be investigated. Given the fact that single-soliton microcombs have facilitated a myriad of applications, ranging from Tb/s coherent communication \cite{marin2017microresonator}, ultrafast ranging \cite{suh2018soliton}, dual-comb spectroscopy \cite{suh2016microresonator}, astronomical spectrometer calibration \cite{obrzud2019microphotonic}, to microwave synthesis \cite{2019arXiv190110372L}, a great benefit for RF filtering can be expected.

\begin{figure*}[t!]
  \centering{
  \includegraphics[width = 1 \linewidth]{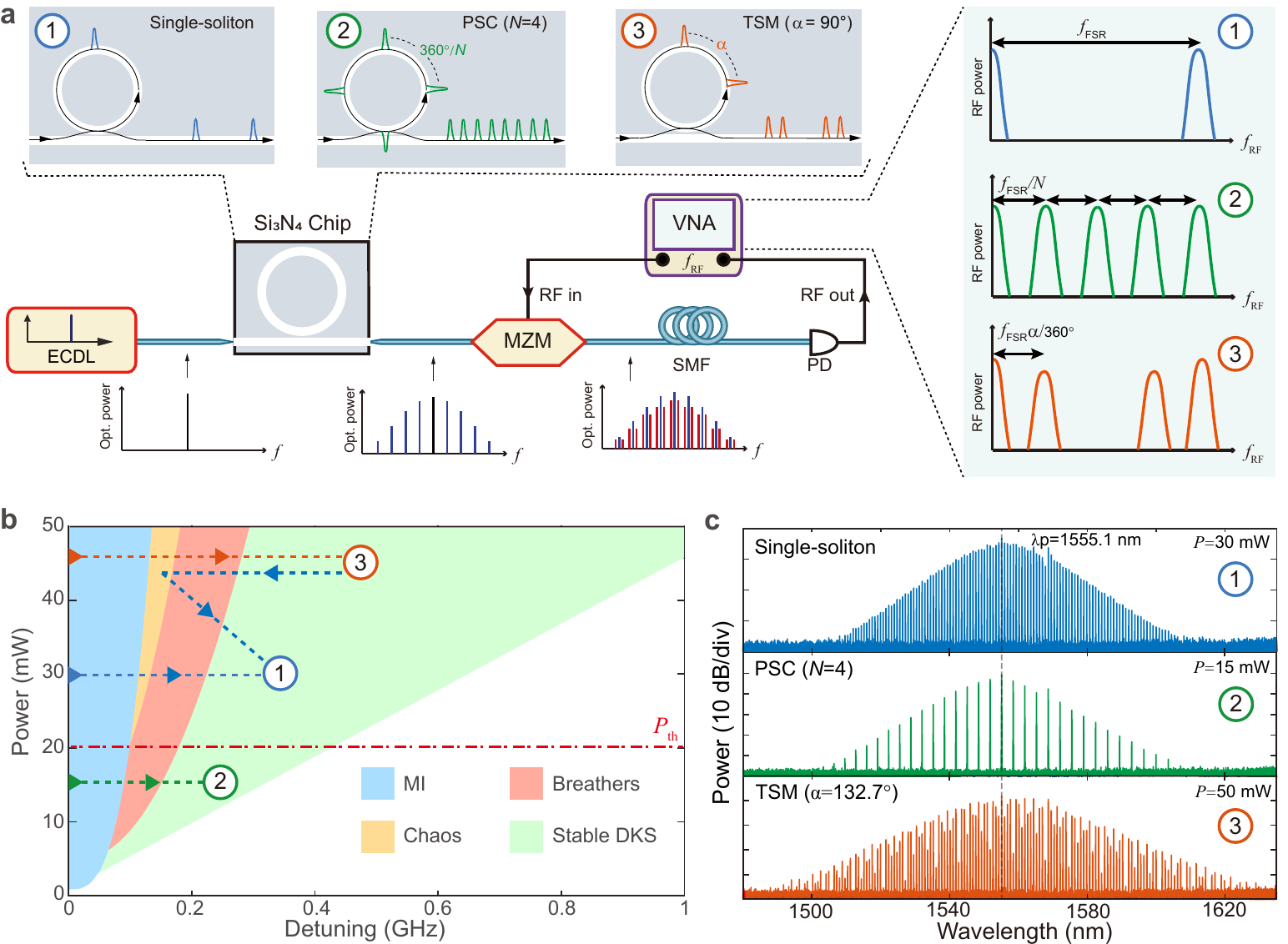}
  }
  \caption{\noindent\textbf{Schematic diagram of reconfigurable soliton based RF photonic filters and their underlying microcomb generation.} \textbf{a,} The conceptual setup consists of four parts: microcomb generation, RF signal upconversion, dispersive propagation, and photodetection. ECDL: external cavity diode laser; MZM: Mach-Zehnder modulator; SMF: single mode fiber; PD: photodiode; VNA: vector network analyzer. Various RF filters are synthesized based on versatile soliton microcombs: (1) single-soliton based RF filter with passband centered at $f_{\mathrm{FSR}}$ (blue); (2) $N-$PSC (perfect soliton crystals of $N$ equally spaced solitons within one round-trip) based RF filters with passband centered at $f_{\mathrm{FSR}}/N$ (green, $N=4$ is shown); (3) Two-soliton microcomb (TSM) based RF filters with passband centered at $f_{\mathrm{FSR}}{\alpha}/360^{\circ}$ (orange), where $\alpha$ is the relative azimuthal angle between two solitons ($\alpha=90^\circ$ is shown). \textbf{b,} Simulated stability diagram of the Lugiato–Lefever equation (LLE) involving the experimental avoided mode crossing (AMX) condition. Four different stability regions are listed: modulation instability (MI, blue), breathers (red), spatio-temporal  and transient chaos (chaos, yellow), and stable dissipative Kerr soliton (DKS, green). PSC and TSM/single-soliton spectra are obtained by distinct approaches. PSC states are accessed under the threshold power to avoid the chaos region. Single-soliton or TSM states are accessed above the threshold power, by either directly falling to the states or backward tuning from higher number of solitons. \textbf{c,} Examples of experimentally generated spectra at resonance of ${1555.1 ~{\rm nm}}$: (1) single-soliton, (2) PSC (N=4), and (3) TSM ($\alpha=132.7^\circ$). The corresponding pump power is also shown for each microcomb generation. 
 \label{fig_setup}
}
\end{figure*} 

In this paper, we demonstrate for the first time, to our knowledge, soliton microcomb based RF photonic filters without any external pulse shaping. In addition, the synthesized RF filters can be all-optically reconfigured through the internal versatile soliton states. Specifically, we trigger, in a deterministic fashion, the perfect soliton crystals (PSC) to multiply the comb spacing \cite{karpov2019dynamics,2019arXiv191000114H}, thereby dividing the RF passband frequencies. Moreover, a completely new regime of filter reconfiguration is achieved based on versatile two-soliton microcombs (TSM). The spectral interference of two solitons is functionally equivalent to an interferometric setup, shifting the filter passband frequency via modification of the angle between them. A proof-of-concept filter reconfiguration experiment is also shown using TSM based RF filters. The internal exploitation of abundant and regulated soliton 
formats of microresonator effectively bypasses the need of another programmable pulse shaper and interferometric setup for RF filter tuning \cite{xue2014programmable,xu2019advanced}. Thus, the proposed scheme dramatically reduces the system complexity and form factor of microcomb based RF filters, and are readily applicable to the current radar systems, 5G wireless, and satellite communications.

Figure \ref{fig_setup}a illustrates the conceptual setup for soliton microcomb based RF filters. Firstly, a telecom C-band continuous wave (CW) laser initiates microcomb generation, where each comb line serves as the RF filter tap. By modulating the RF signals from a vector network analyzer (VNA) on an electro-optic Mach-Zehnder modulator (MZM), the RF signals are broadcast to each microcomb mode. Then the upconverted signals are propagated through a spool of single mode fiber (SMF) to acquire incremental delay between filter taps. Finally, the signals are converted back to RF domain in a fast photodetector (PD). The detailed experimental setup is described in the Methods. This arrangement exactly corresponds to a TDL filter, where the power of each comb line $p_k$ is the tap weight, and the delay is determined by the comb spacing $f_m$ and the accumulated dispersion $ \phi_2 = -{\beta_2}L$ (the product of SMF second-order dispersion $\beta_2$ and fiber length $L$). When the filter tap weights take the ${\rm sech^2}$ envelope of a single-soliton comb (case 1), the RF filter response is given as (see Supplementary Information):
\begin{equation}
H({f_{\mathrm{RF}}})\sim	\cos(2\pi^2{\phi_2}{f_{\mathrm{RF}}^2}) 
\sum_{n=-\infty}^{\infty}G(f_{\mathrm{RF}}-n{f_{\mathrm{FSR}}})
\label{rf_filter_sech}
\end{equation}
where $f_{\mathrm{FSR}}$ and $G(f_{\mathrm{RF}})$ are respectively defined as $1/(2\pi{\phi_2}{f_m})$ and $ 2\frac{T}{T_0}\frac{f_{\mathrm{RF}}}{f_{\mathrm{FSR}}}{\operatorname{sinh}^{-1}
(\frac{T}{T_0}\frac{f_{\mathrm{RF}}}{f_{\mathrm{FSR}}})
}$, with $T=1/f_m$ the repetition period, and $T_0$ the soliton pulse width. $f_{\mathrm{RF}}$ denotes the RF frequency. Notice that higher order dispersion of SMF is neglected here to give a more intuitive picture. The overall RF filter response can be seen as a periodic function of lineshape $ G(f_{\mathrm{RF}})$ with RF free spectral range (FSR) of $f_{\mathrm{FSR}}$, modulated by a envelope due to the double-sideband (DSB) modulation scheme being used. 
As the tap weights are all-positive, the passband frequencies of the RF filters are at every multiples of $f_{\mathrm{FSR}}$, including a DC response. Throughout this letter, we focus on the first passband. 

\begin{figure*}[t!]
  \centering{
 \includegraphics[width = 1 \linewidth]{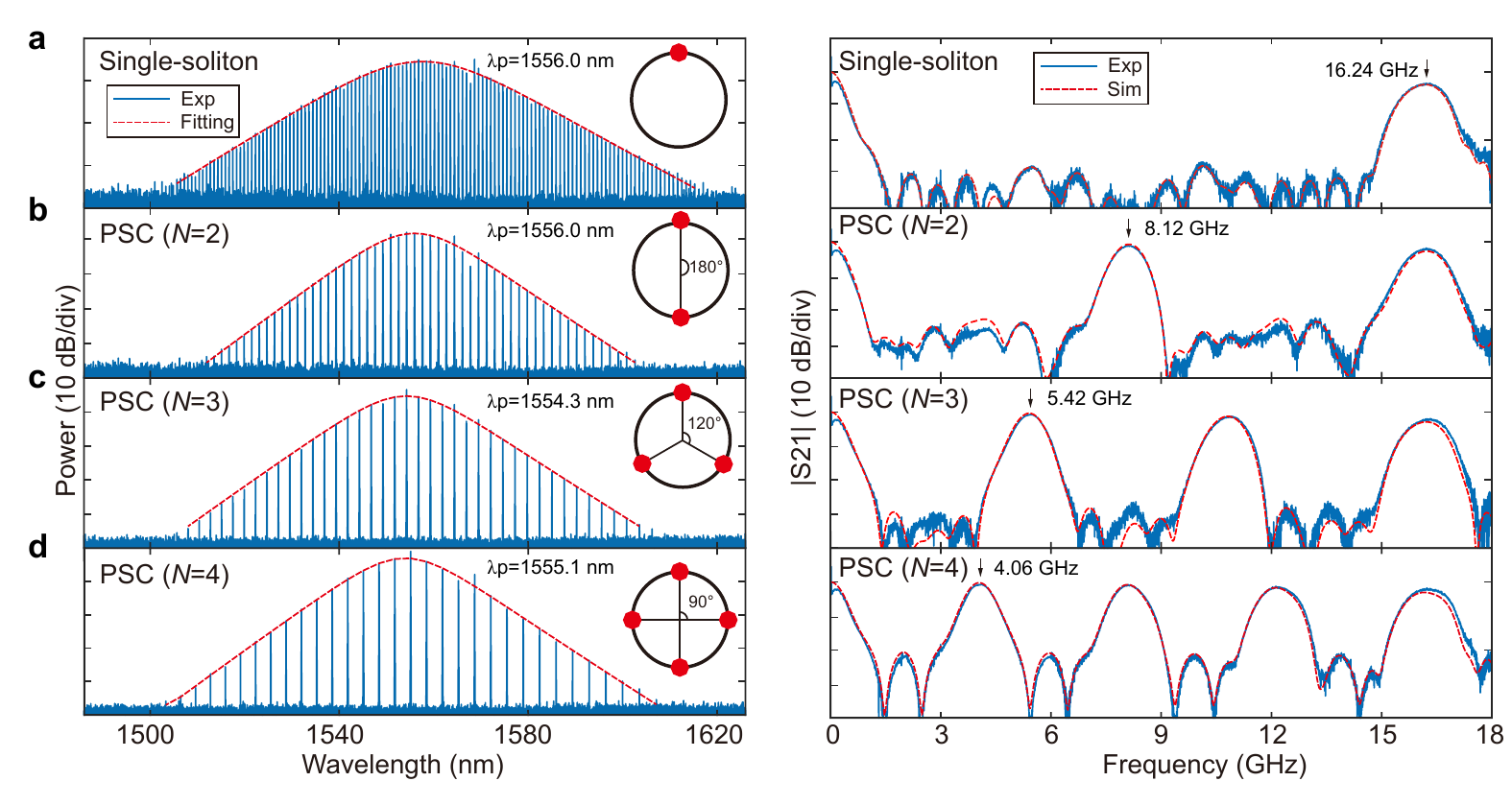}
  }
  \caption{ \noindent \textbf{Single-soliton/PSC spectra and their corresponding RF photonic filters.} Through deterministic accessing PSC states of different resonances, the RF filter passbands can be divided correspondingly. Left column: microcomb spectra (blue: experiment, red: ${\rm sech^2}$ fitting); Right column: corresponding normalized RF filter responses (blue: experiment, red: simulation). \textbf{a,} RF filter centered at ${16.24 ~{\rm GHz}}$ based on single-soliton from resonance of ${1556.0 ~{\rm nm}}$; \textbf{b-d,} RF filters centered at ${8.12 ~{\rm GHz}}$, ${5.42 ~{\rm GHz}}$, and ${4.06 ~{\rm GHz}}$ based on ${2-}$, ${3-}$, ${4-}$times PSC, generated at resonances of ${1556.0 ~{\rm nm}}$, ${1554.3 ~{\rm nm}}$, ${1555.1 ~{\rm nm}}$, respectively. 
 The insets of left column illustrate soliton distribution inside the microresonator: \textbf{a} single-soliton; \textbf{b-d} equally spaced solitons (PSC) with adjacent angles of $180^\circ$, $120^\circ$, and $90^\circ$, respectively $(360^\circ/N, N=2,3,4)$.
 \label{fig_PSC}
}
\end{figure*}

Besides, by exploiting the rich soliton states of microresonator, the RF filters can be easily reconfigured at no additional cost nor complexity. Among soliton crystal structures\cite{cole2017soliton,wang2018robust}, the defect-free PSC is of particular interest, as $ N (N \in N_+ \mid N  \geq 2)$ equally-spaced solitons (case 2) within one roundtrip time simply multiplies the initial comb spacing by $N$-times. This imparts $N$-times division of the filter passband frequencies, while preserving the filter bandwidth. The automatic PSC control is equivalent to the Talbot-based processer for discrete programming the RF filters in ref \cite{maram2019discretely}.
Less intuitively, all-optical reshaping of the RF filters can also be achieved via versatile TSM spectra (case 3). Two solitons residing in one period induce sinusoid interference on the ${\rm sech^2}$ spectral shape of a soliton, modulating the tap weights of the TDL filter. This rewrites the RF filter response as (see Supplementary Information):
\begin{equation}
\begin{aligned}
 H({f_{\mathrm{RF}}}) \sim	\cos(2\pi^2{\phi_2}{f_{\mathrm{RF}}^2}) 
[\sum_{n=-\infty}^{\infty} 2 G(f_{\mathrm{RF}}-n{f_{\mathrm{FSR}}}) \\G(f_{\mathrm{RF}}-(n-\frac{{\alpha}}{2\pi})f_{\mathrm{FSR}})
 + G(f_{\mathrm{RF}}- (n+\frac{{\alpha}}{2\pi})f_{\mathrm{FSR}})]
\label{TSM_response}
\end{aligned}
\end{equation}
where $\alpha$ is the relative azimuthal angle between two solitons (expressed in radian for calculation). Clearly, new RF passbands of halved amplitude appear due to two-soliton interference, which are displaced at both sides from the initial response according to the azimuthal angle between them. Thus, the RF filter passbands can slide inside $f_{\mathrm{FSR}}$ by modifying the relative soliton angles. 
Unlike ref \cite{xu2019advanced} in which the authors artificially introduce the sinusoidal modulation via programming the spectral carving, we alleviate the need for a pulse shaper and realize sinusoidal modulation by directly generating a series of TSM spectra. This novel scheme achieves for the first time wideband reconfiguration of RF filters without either interferometric configuration or additional pulse shaping.

\begin{figure*}[t!]
  \centering{
  \includegraphics[width=1\linewidth]{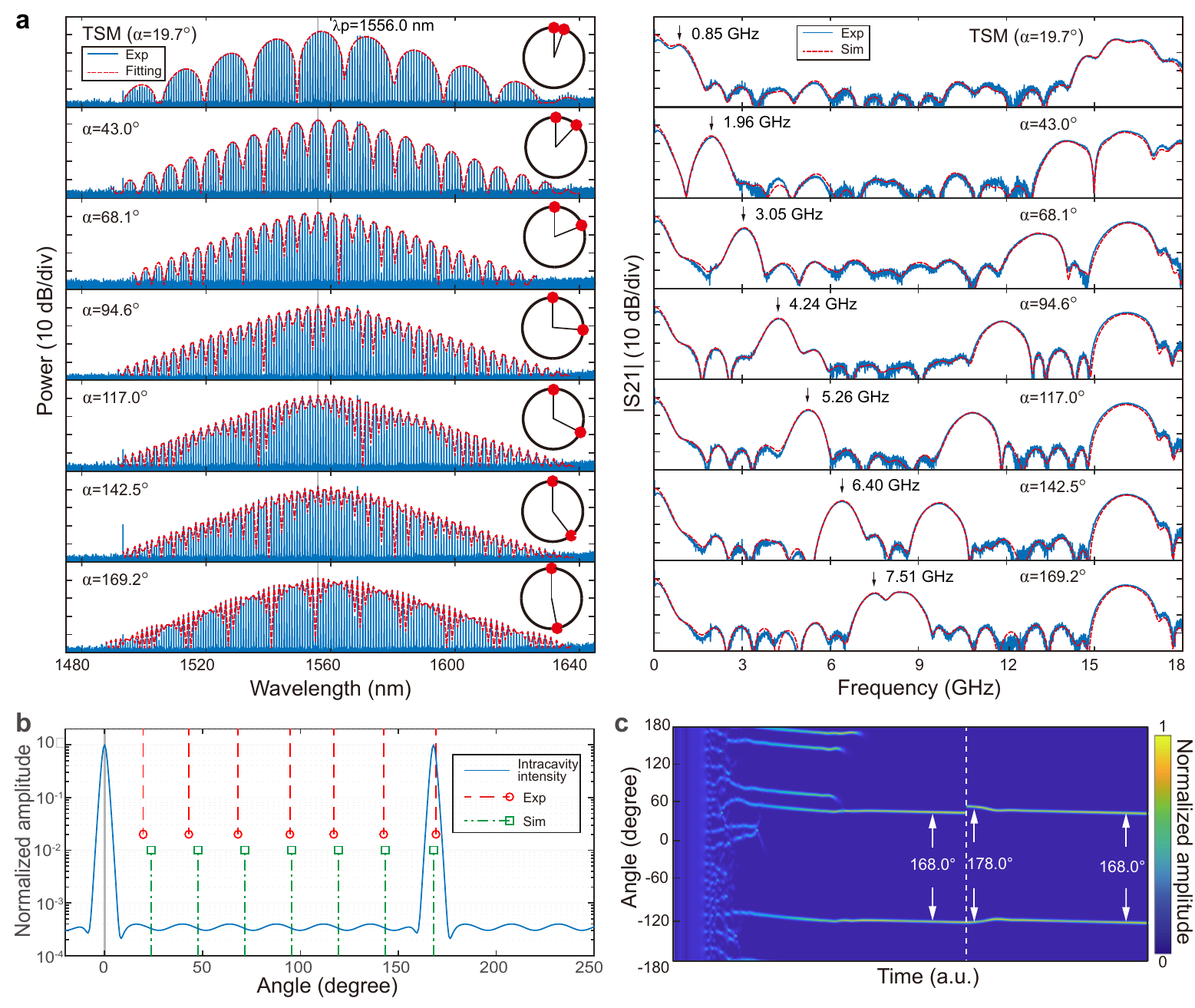}
  }
  \caption{\noindent \textbf{TSM spectra and their corresponding RF photonic filters, together with TSM simulation investigation.} By accessing different two-soliton states, the RF filters can be all-optically reconfigured.
  \textbf{a,} Left column: TSM spectra at resonance of ${1555.96 ~{\rm nm}}$ (blue: experiment, red: envelope fitting). The insets illustrate two soliton distribution inside the microresonator: the angles between them are $ 19.7^\circ$, $ 43.0^\circ$, $ 68.1^\circ$, $ 94.6^\circ$, $ 117.0^\circ$, $ 142.5^\circ$, and $ 169.2^\circ$, respectively. Right column: corresponding normalized RF filter responses (blue: experiment, red: simulation) with passbands at ${0.85}$, ${1.96}$, ${3.05}$, ${4.24}$, ${5.26}$, ${6.40}$, and ${7.51  ~{\rm GHz}}$, respectively. \textbf{b,} Simulation of TSM relative azimuthal angles. One example of the simulated TSM intracavity intensity profile (blue), where AMX induced background modulation is observed. The red and green lines respectively indicate measured and simulated possible relative angles between two solitons.  \textbf{c,} Simulation of the intracavity waveform evolution of TSM for robustness test. First, TSM state with relative angle of $168.0^\circ$ is excited by scanning the pump over the resonance. Once the TSM becomes stable, a $ 10.0^\circ$ perturbation is introduced to one of the solitons at white dashed line. The relative angle will re-stabilize to the original angle of $ 168.0^\circ$ after a period of free running.
  \label{fig_TSM}
  }
\end{figure*}

The soliton microcombs used for RF filtering are generated from an ${104 ~{\rm GHz}}$ ultra-low loss integrated silicon nitride (Si$_3$N$_4$) microresonator (Q $\sim 1\times10^7$), fabricated by the photonic Damascene reflow process \cite{liu2018ultralow}. By employing the frequency-comb-assisted diode laser spectroscopy, the detailed properties of resonances and the integrated group velocity dispersion (GVD) of the microresonator are measured (see Supplementary Information). Strong avoided mode crossings (AMX) are observed around ${1565 ~{\rm nm}}$, which lead to the modulation of intracavity CW background, thereby resulting in the ordering of the DKS pulses \cite{wang2017universal} and the formation of soliton crystals \cite{karpov2019dynamics,cole2017soliton,wang2018robust}. 
Figure \ref{fig_setup}b shows the simulated stability diagram (see Methods), which consists of modulation instability (MI), breathers, chaos (spatio-temporal chaos and transient chaos), and stable DKS states. Additionally, it has been revealed that the pump power level is critical for whether the PSC or stochastic DKS states are formed \cite{karpov2019dynamics}. In our case, the threshold pump power $P_{\mathrm{th}}$ is found to be around $20~{\rm mW}$ in the bus waveguide. When the laser scanning route is operated below threshold pump power, PSC states can be accessed without crossing the chaos region. Contrarily, DKS states with stochastic soliton number are accessed above the threshold power. 
Experimentally, the single soliton and TSM states are obtained by either directing falling to the states or backward tuning from the states with higher soliton number \cite{guo2017universal}. 
Thus, through controlling the pump power and resonance frequency, various soliton microcombs (single-soliton, PSC, and TSM) can be obtained on demand to produce the desired RF filter responses. For example, Figure \ref{fig_setup}c shows three distinct optical spectra obtained from the resonance of ${1555.1 ~{\rm nm}}$: single soliton, PSC ($N=4$), and TSM ($\alpha=132.7^\circ$), respectively.   

Figure \ref{fig_PSC} depicts the RF photonic filters using single soliton and various PSC microcombs. 
The single-soliton based RF filter (Figure \ref{fig_PSC}a) is centered at ${16.24 ~{\rm GHz}}$, with main-to-sidelobe suppression ratio (MSMR) of ${23.2 ~{\rm dB}}$. Further, various PSC states are deterministically obtained at different resonances under the threshold power, thereby all-optically reconfiguring the corresponding RF filters. The comb spacing multiplication via PSC results in the division of the corresponding RF passbands. RF filters centered at ${8.12 ~{\rm GHz}}$, ${5.42 ~{\rm GHz}}$, and ${4.06 ~{\rm GHz}}$ (Figure \ref{fig_PSC}b-d) are experimentally synthesized through $ 2 $, $3$, and $4$ equally spaced solitons, with MSMR of ${22.6 ~{\rm dB}}$, ${25.6 ~{\rm dB}}$, and ${20.4 ~{\rm dB}}$, respectively. All these RF filters achieve MSMR over ${20 ~{\rm dB}}$ without additional programmable spectral shaping. The MSMR here are limited by the smoothness of the optical spectra \cite{supradeepa2012comb}, as several AMX can be seen in the microcombs. Nevertheless, all these microcombs preserve well the $sech^2$ envelope, and remained smooth after amplification. In addition, the measured RF filter responses are in excellent agreement with simulations, by taking into account of third-order dispersion ($\beta_3$) of SMF (see Methods). The bandwidth of the RF filters are respectively ${1060 ~{\rm MHz}}$, ${755 ~{\rm MHz}}$, ${690 ~{\rm MHz}}$, and ${645 ~{\rm MHz}}$, which scale inversely with their center frequencies, also due to the third-order dispersion of SMF \cite{xue2014programmable}.

Figure \ref{fig_TSM}a shows the TSM spectra and their corresponding RF filter responses, pumped at resonance of ${1556.0 ~{\rm nm}}$. According to Eq. \eqref{TSM_response}, the first passband frequencies of RF filters scale linearly with the relative angles between two solitons, so that the filter reconfiguration is achieved. In the experiment, TSM spectra with relative angles of $ 19.7^\circ$, $ 43.0^\circ$, $ 68.1^\circ$, $ 94.6^\circ$, $ 117.0^\circ$, $ 142.5^\circ$, and $ 169.2^\circ$ are obtained, where the angles are extracted from the fitting of the microcomb spectral envelope (see Methods). The measured RF filters are correspondingly centered at ${0.85 ~{\rm GHz}}$, ${1.96 ~{\rm GHz}}$, ${3.05 ~{\rm GHz}}$, ${4.24 ~{\rm GHz}}$, ${5.26 ~{\rm GHz}}$, ${6.40 ~{\rm GHz}}$, and ${7.51 ~{\rm GHz}}$, confirming the linear relation with the soliton angle. 
As in the case of PSC, a slight broadening of the filter passband width from ${490 ~{\rm MHz}}$ to ${620 ~{\rm MHz}}$ is attributed to the third-order dispersion of SMF. Overall, the RF filters obtained at resonance ${1555.96 ~{\rm nm}}$ could vary from DC to ${8.1 ~{\rm GHz}}$ ($f_{\mathrm{FSR}}/2$) with maximum grid of ${1.2 ~{\rm GHz}}$, while roughly preserving the filter bandwidth in the meantime. The granularity of TSM based RF filters can be further reduced to less than ${1 ~{\rm GHz}}$ by exploiting adjacent resonances of ${1556.0 ~{\rm nm}}$ (see Supplementary Information).  



\begin{figure}[t!]
  \centering{
  \includegraphics[width=1\linewidth]{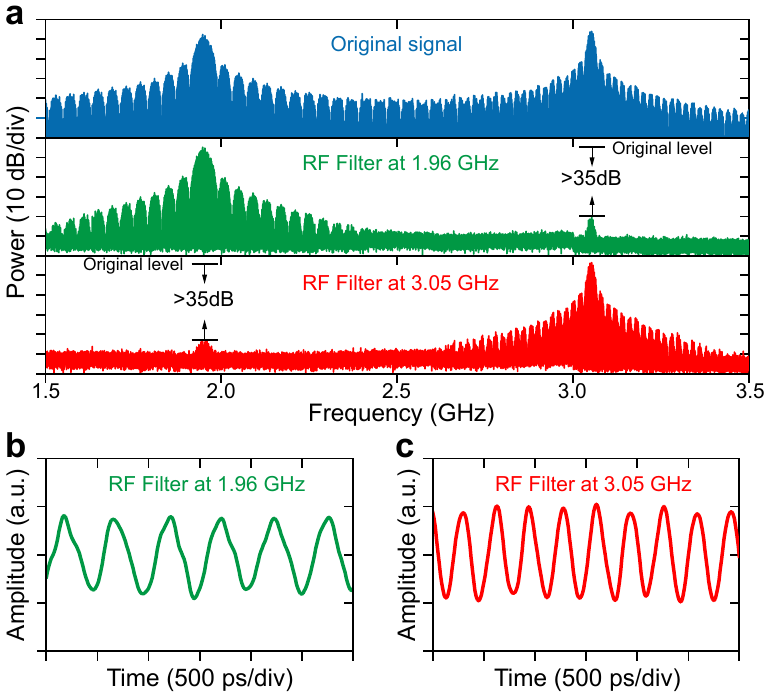}
  }
  \caption{\noindent \textbf{Proof-of-concept experiment using TSM based RF filters.} Two PSK signals with ${40 ~{\rm Mb/s}}$ modulation at ${1.96 ~{\rm GHz}}$ and  ${20 ~{\rm Mb/s}}$ modulation at ${3.05 ~{\rm GHz}}$, are filtered by the TSM based RF filters centered at ${1.96 ~{\rm GHz}}$ and ${3.05 ~{\rm GHz}}$, respectively. \textbf{a} Electrical spectra (from top to bottom) of original signal, signal after ${1.96 ~{\rm GHz}}$ filter, and signal after ${3.05 ~{\rm GHz}}$ filter. \textbf{b} Waveform after ${1.96 ~{\rm GHz}}$ filter. \textbf{c} Waveform after ${3.05 ~{\rm GHz}}$ filter.
    \label{fig_demo}
    }
\end{figure}

Importantly, the possible angles between two solitons are determined by the overall AMX profile, and are rather robust to both laser power and frequency detuning, thereby deterministically dictating the filter passband frequencies to be either one of those shown in Figure \ref{fig_TSM}a. To gain insights of the relative angles between two solitons, we also perform perturbed Lugiato–Lefever equation (LLE) simulation to investigate the TSM formations (see Methods). The blue curve in Figure  \ref{fig_TSM}b shows one example of the steady state two-soliton temporal intracavity profile. Due to the AMX effect, periodic intensity modulation is observed upon the CW background. It is clearly seen that the soliton can only be excited at specific parameter gradients \cite{wang2017universal}, as manifested by the green dashed lines which correspond to the stationary solutions obtained in simulation. These possible soliton angles are in good agreement with experimental results, indicated as red dashed lines. To further test the robustness of the angle between two solitons, an external perturbation is deliberately introduced on their relative angle. Figure \ref{fig_TSM}c illustrates the dynamical evolution of the two-soliton formation. The simulation is initiated as a standard laser scanning scheme to kick out two solitons. Once the simulation reaches stable two-soliton solution (relative angle of $168.0^\circ$), one of the solitons is dragged from its original position by a $10.0^\circ$ on purpose. After a period of free running, the two solitons converge back to their original relative positions, again at $168.0^\circ$ apart. This confirms the regulation of two solitons under AMX background modulation. 

A proof-of-concept RF filter reconfiguration experiment is also illustrated in Figure \ref{fig_demo} using TSM based RF filters (see Methods). Two superimposed phase-shift keying (PSK) signals in which a ${40 ~{\rm Mb/s}}$ modulation at ${1.96 ~{\rm GHz}}$ tone and a ${20 ~{\rm Mb/s}}$ modulation at ${3.05 ~{\rm GHz}}$ tone, are prepared as input test signals. The RF filters are then respectively reconfigured at ${1.96 ~{\rm GHz}}$ and ${3.05 ~{\rm GHz}}$ to filter the input signals, by triggering the TSM spectra of corresponding soliton angles. At the output of the RF filters, nearly complete rejection of either one of the PSK signals is observed on the electrical spectra (Figure \ref{fig_demo}a), where the extinction ratio exceeds ${35 ~{\rm dB}}$ for both cases. Figure \ref{fig_demo}b-c show the filtered output RF waveforms. The periodicity of the output temporal traces corroborate the filtering of the original RF signals. 


 In conclusion, we demonstrate reconfigurable soliton based RF photonic filters using simple approaches. Contrary to previous demonstrations where pulse shapers are necessitate to obtain descent passband responses \cite{xue2014programmable,xu2019advanced,xu2019high}, the proposed schemes are intrinsically well-shaped with the smooth spectral envelope of solitons. More importantly, we harness various intrinsic DKS states of microresonator, like PSC and TSM, for RF filter reconfiguration at no additional cost. The diversity and regularization of soliton formats in microresonator are investigated in the favor of RF photonic filters. To certain extent, these inherent soliton states could be in place of substantial efforts made in the past for reconfiguring the comb based RF filters, such as using interferometric architecture \cite{supradeepa2012comb,xue2014programmable}, programmable pulse shaping \cite{zhu2017novel,xu2019advanced}, or Talbot-based signal processor \cite{maram2019discretely}. Nevertheless, subjected to the same challenges of any other comb based RF filters, our current filters are not yet optimized in terms of the link performance. While the noise reduction and gain enhancement could be achieved using high power-handling balanced detectors \cite{kim2014comb}.
 Besides, the recent advancement on the integration between laser chip and microresonator \cite{stern2018battery,raja2019electrically}, as well as replacing the SMF with a highly dispersive integrated waveguide \cite{sancho2012integrable}, can be further connected to the current work for miniaturization. To conclude, our work significantly reduces the system complexity, size, and cost of the microcomb based RF filters, while preserving their wide reconfigurability. The proposed schemes set as a stepping stone for chipscale, cost-effective, and widely reconfigurable microcomb based RF filters. 

\vspace{0.5cm}

\noindent\textbf{Methods}
\medskip
\begin{footnotesize}

\noindent \textbf{Experimental setup}:
A C-band tunable CW laser is amplified by an Erbium-doped fiber amplifier (EDFA) with amplified spontaneous emission (ASE) filtered, polarization aligned at the TE mode, and then coupled to the ${\rm Si_3N_4}$ microresonator for soliton microcomb generation. The input and output coupling of the chip is achieved via lensed fibers of around ${30 \%}$ fiber-chip-fiber coupling efficiency. The soliton microcombs are initiated by scanning the pump over the resonances, with the assistance of an arbitrary function generator (AFG) \cite{herr2014temporal}. The residual pump of generated microcombs are then filtered by a tunable fiber Bragg grating (FBG), while a circulator is inserted in between to avoid back-reflection. ${10 \%}$ of light is tapped to an optical spectrum analyzer (OSA) to record the microcomb spectra. The other ${90 \%}$ of the light is amplified, and polarization managed, before sending to a ${30 ~{\rm GHz}}$ bandiwdth MZM. RF signals from the VNA are applied to the MZM in DSB modulation format. The modulated spectra are then propagated through a spool of ${4583.8 ~{\rm m}}$ SMF to acquire dispersive delays, and finally beats at a ${18 ~{\rm GHz}}$ PD to convert the signals back to the RF domain. The length of SMF is measured by a commercial optical time-domain reflectometer (OTDR).

For the system demonstration, a ${12 ~{\rm GSa/s}}$ arbitrary waveform generator (AWG) is used to prepare the input RF signals. ${40 ~{\rm Mb/s}}$ PSK signal modulated at ${1.96 ~{\rm GHz}}$ tone and ${20 ~{\rm Mb/s}}$ PSK signal modulated at ${3.05 ~{\rm GHz}}$ tone, are generated separately from the two channels of the AWG. After adding the two streams of signals in a combiner, the composite signal is then sent through the TSM based RF filters, tuned at ${1.96 ~{\rm GHz}}$ and ${3.05 ~{\rm GHz}}$, respectively. The output spectra are measured by a electrical spectrum analyzer (ESA), while the waveforms are measured using a high-speed real-time oscilloscope. 

\noindent \textbf{Si$_3$N$_4$ microresonator}:
The Si$_3$N$_4$ microresonator used in the experiment is a ring structure with radius of $ 217 ~{\rm \mu m}$. Its waveguide cross section (width $\times$ height), is made to be $ 1500 ~{\rm nm}\times750 ~{\rm nm}$. The microresonator is coupled with a bus waveguide, which possesses the same cross section as the ring to realize high coupling ideality\cite{liu2018ultralow}. To achieve critical coupling for the resonances, the gap distance between the ring and bus waveguide is designed to be $ 690 ~{\rm nm}$. In our experiment, the pumped resonances are around $ 1556 ~{\rm nm}$, where both the intrinsic linewidths and coupling strengths are approximately $20 ~{\rm MHz}$ (see Supplementary Information). With respect to the reference resonance of $\omega_0/2\pi=192.8 ~{\rm THz}$, the dispersion parameters of microresonator are measured: FSR of microresonator $D_1/2\pi\approx103.9 ~{\rm GHz}$, second-order dispersion term $D_2/2\pi\approx1.28 ~{\rm MHz}$, and negligible third-order dispersion term $D_3/2\pi\sim\mathcal{O}(1)~{\rm kHz}$ (see Supplementary Information).

\noindent \textbf{LLE Simulation}:
The simulation performed in this work is based on the perturbed LLE model: 
\begin{equation}
\begin{aligned} 
\frac{\partial A(\phi ,t)}{\partial t}= & -(\frac{\kappa }{2}+j(\omega _0-\omega _p))A(\phi,t)+j\frac{D_2}{2}\frac{\partial^2 A(\phi ,t)}{\partial \phi ^2}\\
& +jg|A(\phi ,t)|^2A(\phi ,t)+ \sqrt{\kappa_{\mathrm{ex}}}s_{\mathrm{in}}
\label{LLE}
\end{aligned}
\end{equation}
where $A(\phi,t)$ is the temporal envelope of the intracavity field. $\kappa=\kappa_{\mathrm{ex}}+\kappa_0$ is the total cavity loss rate, where $\kappa_{\mathrm{ex}}$ is the coupling rate, and $\kappa_0$ is the internal loss rate. 
$\omega _0$ and $\omega _p$ denote the angular frequencies of the pumped resonance and the CW pump laser, respectively. 
$g$ is the Kerr frequency shift per photon, defined as $g=\hbar\omega_0^2cn_2/n_0^2V_{\mathrm{eff}}$, where $n_0$ is the effective group refractive index, $n_2$ is the nonlinear optical index, and $V_{\mathrm{eff}}$ is the effective mode volume.  $D_2$ corresponds to the second-order dispersion term, and $|s_{\mathrm{in}}|^2$ is the pump power.
To involve the AMX effect, an additional frequency detuning $\Delta_k$ is introduced at the $k$-th mode, so that the frequency of the $k$-th mode becomes $\omega_k=\omega_0+D_1k+D_2k^2/2+\Delta_k$ (see Supplementary Information). 
Here in simulation, the dispersion is limited to $D_2$, and the Raman and thermal effects are not taken into account.
According to the dispersion measurement and the generated microcomb spectra, the parameters for the AMX in the simulation are set as $k=15$ and $\Delta_k/2\pi=100~\mathrm{MHz}$, enabling the modulation of the CW intracavity background for the trapping of soliton temporal positions. Note that the strength of the AMX here is estimated to introduce the regularizability of solitons, but without disturbing their formations\cite{karpov2019dynamics}. Other parameters used in the simulation are retrieved from the characterization, that is, $D_1/2\pi=103.9~\mathrm{GHz}$, $D_2/2\pi=1.28~\mathrm{MHz}$, $\kappa_{\mathrm{ex}}/{2\pi} = \kappa_0/{2\pi} = \kappa/{4\pi} = 20~\mathrm{MHz}$.

The simulation of stability chart is obtained by initializing the numerical model with single-soliton solution at various pump power and detuning conditions \cite{karpov2019dynamics}. Four different states are found: MI, breathers, spatiotemporal and transient chaos, and stable DKS states. The threshold pump power, separating the PSC and stochastic DKS formations, is estimated from both the simulation and experimental results.
For the TSM simulation, the numerical model is operated under standard CW laser pump scanning from blue-detuned to red-detuned side, until it reaches the stable TSM states. All the possible angles of TSM are recorded. To test the robustness of the TSM azimuthal angle, the model is initialized with one of the exact two-soliton solution but deliberately perturbed by $10^\circ$ angle deviation. Two solitons are gradually re-stabilized at its original azimuthal angle after a period of free runing.

\noindent \textbf{RF filter response fitting}:
As the generated microcomb spectra are broader than the amplifying bandwidth of EDFA, we also measured the optical spectra after the EDFA, in order to extract the TDL filter tap weights. The third-order dispersion $\beta_3$ of SMF is taken into account for the fitting of RF responses, which can be formulated as \cite{zhu2017novel}:
\begin{equation}
\begin{aligned} 
& H({f_{\mathrm{RF}}})\sim	
\sum_{k}^{} {p_{k}}\cos(2\pi^2{\phi_2}{f_{\mathrm{RF}}^2} +  4\pi^3{\phi_3}k{f_m^2}{f_{\mathrm{RF}}}) \times \\
& \exp(j{4\pi^2}{\phi_2}k{f_m}{f_{\mathrm{RF}}} + j{4\pi^3}{\phi_3}{k^2}{f_m^2}{f_{\mathrm{RF}}} + j\frac{4}{3}{\pi^3}{\phi_3}{f_m^3}) \\
\end{aligned} 
\end{equation}
where $\phi_3=-{\beta_3}L$. In accordance with typical values of SMF dispersion, $ {\beta_2}= - 20.2 ~{\rm ps^2/km}$ and $ {\beta_3}= 0.117 ~{\rm ps^3/km}$ at $ 1550 ~{\rm nm}$ are estimated for all the above fittings of RF filters. The simulation results are in excellent agreement with experimental RF filter responses.

\noindent \textbf{TSM spectral fitting}:
First, we extract the power of each comb mode of experimental TSM spectra, and indexed them with respect to the pump mode. Pump mode is rejected and amplitude rescaling is considered as a fitting parameter. Note that the amount of spectral red-shift due to Raman effect is also estimated in fitting, by displacing the center of ${\rm sech^2}$ soliton spectra from the pump comb mode. Then, the rescaling parameter and red-shift are estimated to best fit the experimental data with the TSM spectral power equation (see Supplementary Information), thereby retrieving the azimthual angle $\alpha$ between two solitons. Excellent match between simulations and experimental spectra are obtained. 

\noindent \textbf{Funding Information}:
This work was supported by Contract HR0011-15-C-0055 (DODOS) from the Defense Advanced Research Projects Agency (DARPA), Microsystems Technology Office (MTO), and by the Air Force Office of Scientific Research, Air Force Material Command, USAF under Award No. FA9550-19-1-0250, and by Swiss National Science Foundation under grant agreement No. 176563 (BRIDGE), No.165933 and No. 159897.

\noindent \textbf{Acknowledgments}: 
The Si$_3$N$_4$ microresonator samples were fabricated in the EPFL center of MicroNanoTechnology (CMi).

\noindent \textbf{Data Availability Statement}: 
The code and data used to produce the plots within this work will be released on the repository \texttt{Zenodo} upon publication of this preprint.
\end{footnotesize}


\bibliographystyle{naturemag}
\bibliography{ref}

\begin{thebibliography}{10}
\expandafter\ifx\csname url\endcsname\relax
  \def\url#1{\texttt{#1}}\fi
\expandafter\ifx\csname urlprefix\endcsname\relax\def\urlprefix{URL }\fi
\providecommand{\bibinfo}[2]{#2}
\providecommand{\eprint}[2][]{\url{#2}}

\bibitem{xue2014programmable}
\bibinfo{author}{Xue, X.} \emph{et~al.}
\newblock \bibinfo{title}{Programmable single-bandpass photonic rf filter based
  on kerr comb from a microring}.
\newblock \emph{\bibinfo{journal}{Journal of Lightwave Technology}}
  \textbf{\bibinfo{volume}{32}}, \bibinfo{pages}{3557--3565}
  (\bibinfo{year}{2014}).
\newblock \urlprefix\url{https://doi.org/10.1109/jlt.2014.2312359}.

\bibitem{xu2019advanced}
\bibinfo{author}{Xu, X.} \emph{et~al.}
\newblock \bibinfo{title}{Advanced adaptive photonic rf filters with 80 taps
  based on an integrated optical micro-comb source}.
\newblock \emph{\bibinfo{journal}{Journal of Lightwave Technology}}
  \textbf{\bibinfo{volume}{37}}, \bibinfo{pages}{1288--1295}
  (\bibinfo{year}{2019}).
\newblock \urlprefix\url{https://doi.org/10.1109/jlt.2019.2892158}.

\bibitem{xu2019high}
\bibinfo{author}{Xu, X.} \emph{et~al.}
\newblock \bibinfo{title}{High performance rf filters via bandwidth scaling
  with kerr micro-combs}.
\newblock \emph{\bibinfo{journal}{APL Photonics}} \textbf{\bibinfo{volume}{4}},
  \bibinfo{pages}{026102} (\bibinfo{year}{2019}).
\newblock \urlprefix\url{https://doi.org/10.1063/1.5080246}.

\bibitem{karpov2019dynamics}
\bibinfo{author}{Karpov, M.} \emph{et~al.}
\newblock \bibinfo{title}{Dynamics of soliton crystals in optical
  microresonators}.
\newblock \emph{\bibinfo{journal}{Nature Physics}} \bibinfo{pages}{1--7}
  (\bibinfo{year}{2019}).
\newblock \urlprefix\url{https://doi.org/10.1038/s41567-019-0635-0}.

\bibitem{wang2017universal}
\bibinfo{author}{Wang, Y.} \emph{et~al.}
\newblock \bibinfo{title}{Universal mechanism for the binding of temporal
  cavity solitons}.
\newblock \emph{\bibinfo{journal}{Optica}} \textbf{\bibinfo{volume}{4}},
  \bibinfo{pages}{855--863} (\bibinfo{year}{2017}).
\newblock \urlprefix\url{https://doi.org/10.1364/optica.4.000855}.

\bibitem{capmany2007microwave}
\bibinfo{author}{Capmany, J.} \& \bibinfo{author}{Novak, D.}
\newblock \bibinfo{title}{Microwave photonics combines two worlds}.
\newblock \emph{\bibinfo{journal}{Nature photonics}}
  \textbf{\bibinfo{volume}{1}}, \bibinfo{pages}{319} (\bibinfo{year}{2007}).
\newblock \urlprefix\url{https://doi.org/10.1038/nphoton.2007.89}.

\bibitem{marpaung2019integrated}
\bibinfo{author}{Marpaung, D.}, \bibinfo{author}{Yao, J.} \&
  \bibinfo{author}{Capmany, J.}
\newblock \bibinfo{title}{Integrated microwave photonics}.
\newblock \emph{\bibinfo{journal}{Nature photonics}}
  \textbf{\bibinfo{volume}{13}}, \bibinfo{pages}{80} (\bibinfo{year}{2019}).
\newblock \urlprefix\url{https://doi.org/10.1038/s41566-018-0310-5}.

\bibitem{sancho2012integrable}
\bibinfo{author}{Sancho, J.} \emph{et~al.}
\newblock \bibinfo{title}{Integrable microwave filter based on a photonic
  crystal delay line}.
\newblock \emph{\bibinfo{journal}{Nature communications}}
  \textbf{\bibinfo{volume}{3}}, \bibinfo{pages}{1075} (\bibinfo{year}{2012}).
\newblock \urlprefix\url{https://doi.org/10.1038/ncomms2092}.

\bibitem{metcalf2016integrated}
\bibinfo{author}{Metcalf, A.~J.} \emph{et~al.}
\newblock \bibinfo{title}{Integrated line-by-line optical pulse shaper for
  high-fidelity and rapidly reconfigurable rf-filtering}.
\newblock \emph{\bibinfo{journal}{Optics express}}
  \textbf{\bibinfo{volume}{24}}, \bibinfo{pages}{23925--23940}
  (\bibinfo{year}{2016}).
\newblock \urlprefix\url{https://doi.org/10.1364/oe.24.023925}.

\bibitem{zhuang2015programmable}
\bibinfo{author}{Zhuang, L.}, \bibinfo{author}{Roeloffzen, C.~G.},
  \bibinfo{author}{Hoekman, M.}, \bibinfo{author}{Boller, K.-J.} \&
  \bibinfo{author}{Lowery, A.~J.}
\newblock \bibinfo{title}{Programmable photonic signal processor chip for
  radiofrequency applications}.
\newblock \emph{\bibinfo{journal}{Optica}} \textbf{\bibinfo{volume}{2}},
  \bibinfo{pages}{854--859} (\bibinfo{year}{2015}).
\newblock \urlprefix\url{https://doi.org/10.1364/optica.2.000854}.

\bibitem{marpaung2013si}
\bibinfo{author}{Marpaung, D.} \emph{et~al.}
\newblock \bibinfo{title}{Si 3 n 4 ring resonator-based microwave photonic
  notch filter with an ultrahigh peak rejection}.
\newblock \emph{\bibinfo{journal}{Optics express}}
  \textbf{\bibinfo{volume}{21}}, \bibinfo{pages}{23286--23294}
  (\bibinfo{year}{2013}).
\newblock \urlprefix\url{https://doi.org/10.1364/oe.21.023286}.

\bibitem{eggleton2019brillouin}
\bibinfo{author}{Eggleton, B.~J.}, \bibinfo{author}{Poulton, C.~G.},
  \bibinfo{author}{Rakich, P.~T.}, \bibinfo{author}{Steel, M.~J.} \&
  \bibinfo{author}{Bahl, G.}
\newblock \bibinfo{title}{Brillouin integrated photonics}.
\newblock \emph{\bibinfo{journal}{Nature Photonics}} \bibinfo{pages}{1--14}
  (\bibinfo{year}{2019}).
\newblock \urlprefix\url{https://doi.org/10.1038/s41566-019-0498-z}.

\bibitem{fandino2017monolithic}
\bibinfo{author}{Fandi{\~n}o, J.~S.}, \bibinfo{author}{Mu{\~n}oz, P.},
  \bibinfo{author}{Dom{\'e}nech, D.} \& \bibinfo{author}{Capmany, J.}
\newblock \bibinfo{title}{A monolithic integrated photonic microwave filter}.
\newblock \emph{\bibinfo{journal}{Nature Photonics}}
  \textbf{\bibinfo{volume}{11}}, \bibinfo{pages}{124} (\bibinfo{year}{2017}).
\newblock \urlprefix\url{https://doi.org/10.1038/nphoton.2016.233}.

\bibitem{capmany2006tutorial}
\bibinfo{author}{Capmany, J.}, \bibinfo{author}{Ortega, B.} \&
  \bibinfo{author}{Pastor, D.}
\newblock \bibinfo{title}{A tutorial on microwave photonic filters}.
\newblock \emph{\bibinfo{journal}{Journal of Lightwave Technology}}
  \textbf{\bibinfo{volume}{24}}, \bibinfo{pages}{201--229}
  (\bibinfo{year}{2006}).
\newblock \urlprefix\url{https://doi.org/10.1109/jlt.2005.860478}.

\bibitem{supradeepa2012comb}
\bibinfo{author}{Supradeepa, V.} \emph{et~al.}
\newblock \bibinfo{title}{Comb-based radiofrequency photonic filters with rapid
  tunability and high selectivity}.
\newblock \emph{\bibinfo{journal}{Nature Photonics}}
  \textbf{\bibinfo{volume}{6}}, \bibinfo{pages}{186} (\bibinfo{year}{2012}).
\newblock \urlprefix\url{https://doi.org/10.1038/nphoton.2011.350}.

\bibitem{maram2019discretely}
\bibinfo{author}{Maram, R.}, \bibinfo{author}{Onori, D.},
  \bibinfo{author}{Aza{\~n}a, J.} \& \bibinfo{author}{Chen, L.~R.}
\newblock \bibinfo{title}{Discretely programmable microwave photonic filter
  based on temporal talbot effects}.
\newblock \emph{\bibinfo{journal}{Optics express}}
  \textbf{\bibinfo{volume}{27}}, \bibinfo{pages}{14381--14391}
  (\bibinfo{year}{2019}).
\newblock \urlprefix\url{https://doi.org/10.1364/oe.27.014381}.

\bibitem{zhu2017novel}
\bibinfo{author}{Zhu, X.}, \bibinfo{author}{Chen, F.}, \bibinfo{author}{Peng,
  H.} \& \bibinfo{author}{Chen, Z.}
\newblock \bibinfo{title}{Novel programmable microwave photonic filter with
  arbitrary filtering shape and linear phase}.
\newblock \emph{\bibinfo{journal}{Optics express}}
  \textbf{\bibinfo{volume}{25}}, \bibinfo{pages}{9232--9243}
  (\bibinfo{year}{2017}).
\newblock \urlprefix\url{https://doi.org/10.1364/oe.25.009232}.

\bibitem{xue2018microcomb}
\bibinfo{author}{Xue, X.} \emph{et~al.}
\newblock \bibinfo{title}{Microcomb-based true-time-delay network for microwave
  beamforming with arbitrary beam pattern control}.
\newblock \emph{\bibinfo{journal}{Journal of Lightwave Technology}}
  \textbf{\bibinfo{volume}{36}}, \bibinfo{pages}{2312--2321}
  (\bibinfo{year}{2018}).
\newblock \urlprefix\url{https://doi.org/10.1109/jlt.2018.2803743}.

\bibitem{xu2018broadband}
\bibinfo{author}{Xu, X.} \emph{et~al.}
\newblock \bibinfo{title}{Broadband rf channelizer based on an integrated
  optical frequency kerr comb source}.
\newblock \emph{\bibinfo{journal}{Journal of Lightwave Technology}}
  \textbf{\bibinfo{volume}{36}}, \bibinfo{pages}{4519--4526}
  (\bibinfo{year}{2018}).
\newblock \urlprefix\url{https://doi.org/10.1109/jlt.2018.2819172}.

\bibitem{tan2019microwave}
\bibinfo{author}{Tan, M.} \emph{et~al.}
\newblock \bibinfo{title}{Microwave and rf photonic fractional hilbert
  transformer based on a 50ghz kerr micro-comb}.
\newblock \emph{\bibinfo{journal}{Journal of Lightwave Technology}}
  (\bibinfo{year}{2019}).
\newblock \urlprefix\url{https://doi.org/10.1109/jlt.2019.2946606}.

\bibitem{xue2015mode}
\bibinfo{author}{Xue, X.} \emph{et~al.}
\newblock \bibinfo{title}{Mode-locked dark pulse kerr combs in
  normal-dispersion microresonators}.
\newblock \emph{\bibinfo{journal}{Nature Photonics}}
  \textbf{\bibinfo{volume}{9}}, \bibinfo{pages}{594} (\bibinfo{year}{2015}).
\newblock \urlprefix\url{https://doi.org/10.1038/nphoton.2015.137}.

\bibitem{herr2014temporal}
\bibinfo{author}{Herr, T.} \emph{et~al.}
\newblock \bibinfo{title}{Temporal solitons in optical microresonators}.
\newblock \emph{\bibinfo{journal}{Nature Photonics}}
  \textbf{\bibinfo{volume}{8}}, \bibinfo{pages}{145} (\bibinfo{year}{2014}).
\newblock \urlprefix\url{https://doi.org/10.1038/nphoton.2013.343}.

\bibitem{stern2018battery}
\bibinfo{author}{Stern, B.}, \bibinfo{author}{Ji, X.},
  \bibinfo{author}{Okawachi, Y.}, \bibinfo{author}{Gaeta, A.~L.} \&
  \bibinfo{author}{Lipson, M.}
\newblock \bibinfo{title}{Battery-operated integrated frequency comb
  generator}.
\newblock \emph{\bibinfo{journal}{Nature}} \textbf{\bibinfo{volume}{562}},
  \bibinfo{pages}{401} (\bibinfo{year}{2018}).
\newblock \urlprefix\url{https://doi.org/10.1038/s41586-018-0598-9}.

\bibitem{raja2019electrically}
\bibinfo{author}{Raja, A.~S.} \emph{et~al.}
\newblock \bibinfo{title}{Electrically pumped photonic integrated soliton
  microcomb}.
\newblock \emph{\bibinfo{journal}{Nature communications}}
  \textbf{\bibinfo{volume}{10}}, \bibinfo{pages}{680} (\bibinfo{year}{2019}).
\newblock \urlprefix\url{https://doi.org/10.1038/s41467-019-08498-2}.

\bibitem{marin2017microresonator}
\bibinfo{author}{Marin-Palomo, P.} \emph{et~al.}
\newblock \bibinfo{title}{Microresonator-based solitons for massively parallel
  coherent optical communications}.
\newblock \emph{\bibinfo{journal}{Nature}} \textbf{\bibinfo{volume}{546}},
  \bibinfo{pages}{274} (\bibinfo{year}{2017}).
\newblock \urlprefix\url{https://doi.org/10.1038/nature22387}.

\bibitem{suh2018soliton}
\bibinfo{author}{Suh, M.-G.} \& \bibinfo{author}{Vahala, K.~J.}
\newblock \bibinfo{title}{Soliton microcomb range measurement}.
\newblock \emph{\bibinfo{journal}{Science}} \textbf{\bibinfo{volume}{359}},
  \bibinfo{pages}{884--887} (\bibinfo{year}{2018}).
\newblock \urlprefix\url{https://doi.org/10.1126/science.aao1968}.

\bibitem{suh2016microresonator}
\bibinfo{author}{Suh, M.-G.}, \bibinfo{author}{Yang, Q.-F.},
  \bibinfo{author}{Yang, K.~Y.}, \bibinfo{author}{Yi, X.} \&
  \bibinfo{author}{Vahala, K.~J.}
\newblock \bibinfo{title}{Microresonator soliton dual-comb spectroscopy}.
\newblock \emph{\bibinfo{journal}{Science}} \textbf{\bibinfo{volume}{354}},
  \bibinfo{pages}{600--603} (\bibinfo{year}{2016}).
\newblock \urlprefix\url{https://doi.org/10.1126/science.aah6516}.

\bibitem{obrzud2019microphotonic}
\bibinfo{author}{Obrzud, E.} \emph{et~al.}
\newblock \bibinfo{title}{A microphotonic astrocomb}.
\newblock \emph{\bibinfo{journal}{Nature Photonics}}
  \textbf{\bibinfo{volume}{13}}, \bibinfo{pages}{31} (\bibinfo{year}{2019}).
\newblock \urlprefix\url{https://doi.org/10.1038/s41566-018-0309-y}.

\bibitem{2019arXiv190110372L}
\bibinfo{author}{{Liu}, J.} \emph{et~al.}
\newblock \bibinfo{title}{{Nanophotonic soliton-based microwave synthesizers}}.
\newblock \emph{\bibinfo{journal}{arXiv e-prints}}
  \bibinfo{pages}{arXiv:1901.10372} (\bibinfo{year}{2019}).
\newblock \urlprefix\url{https://arxiv.org/abs/1901.10372}.

\bibitem{2019arXiv191000114H}
\bibinfo{author}{{He}, Y.}, \bibinfo{author}{{Ling}, J.},
  \bibinfo{author}{{Li}, M.} \& \bibinfo{author}{{Lin}, Q.}
\newblock \bibinfo{title}{{Perfect soliton crystals on demand}}.
\newblock \emph{\bibinfo{journal}{arXiv e-prints}}
  \bibinfo{pages}{arXiv:1910.00114} (\bibinfo{year}{2019}).
\newblock \urlprefix\url{https://arxiv.org/abs/1910.00114}.

\bibitem{cole2017soliton}
\bibinfo{author}{Cole, D.~C.}, \bibinfo{author}{Lamb, E.~S.},
  \bibinfo{author}{Del’Haye, P.}, \bibinfo{author}{Diddams, S.~A.} \&
  \bibinfo{author}{Papp, S.~B.}
\newblock \bibinfo{title}{Soliton crystals in kerr resonators}.
\newblock \emph{\bibinfo{journal}{Nature Photonics}}
  \textbf{\bibinfo{volume}{11}}, \bibinfo{pages}{671} (\bibinfo{year}{2017}).
\newblock \urlprefix\url{https://doi.org/10.1038/s41566-017-0009-z}.

\bibitem{wang2018robust}
\bibinfo{author}{Wang, W.} \emph{et~al.}
\newblock \bibinfo{title}{Robust soliton crystals in a thermally controlled
  microresonator}.
\newblock \emph{\bibinfo{journal}{Optics letters}}
  \textbf{\bibinfo{volume}{43}}, \bibinfo{pages}{2002--2005}
  (\bibinfo{year}{2018}).
\newblock \urlprefix\url{https://doi.org/10.1364/ol.43.002002}.

\bibitem{liu2018ultralow}
\bibinfo{author}{Liu, J.} \emph{et~al.}
\newblock \bibinfo{title}{Ultralow-power chip-based soliton microcombs for
  photonic integration}.
\newblock \emph{\bibinfo{journal}{Optica}} \textbf{\bibinfo{volume}{5}},
  \bibinfo{pages}{1347--1353} (\bibinfo{year}{2018}).
\newblock \urlprefix\url{https://doi.org/10.1364/optica.5.001347}.

\bibitem{guo2017universal}
\bibinfo{author}{Guo, H.} \emph{et~al.}
\newblock \bibinfo{title}{Universal dynamics and deterministic switching of
  dissipative kerr solitons in optical microresonators}.
\newblock \emph{\bibinfo{journal}{Nature Physics}}
  \textbf{\bibinfo{volume}{13}}, \bibinfo{pages}{94} (\bibinfo{year}{2017}).
\newblock \urlprefix\url{https://doi.org/10.1038/nphys3893}.

\bibitem{kim2014comb}
\bibinfo{author}{Kim, H.-J.}, \bibinfo{author}{Leaird, D.~E.},
  \bibinfo{author}{Metcalf, A.~J.} \& \bibinfo{author}{Weiner, A.~M.}
\newblock \bibinfo{title}{Comb-based rf photonic filters based on
  interferometric configuration and balanced detection}.
\newblock \emph{\bibinfo{journal}{Journal of Lightwave Technology}}
  \textbf{\bibinfo{volume}{32}}, \bibinfo{pages}{3478--3488}
  (\bibinfo{year}{2014}).
\newblock \urlprefix\url{https://doi.org/10.1109/jlt.2014.2326410}.

\end{thebibliography}


\begin{thebibliography}{1}
\expandafter\ifx\csname url\endcsname\relax
  \def\url#1{\texttt{#1}}\fi
\expandafter\ifx\csname urlprefix\endcsname\relax\def\urlprefix{URL }\fi
\providecommand{\bibinfo}[2]{#2}
\providecommand{\eprint}[2][]{\url{#2}}

\bibitem{zhu2017novel}
\bibinfo{author}{Zhu, X.}, \bibinfo{author}{Chen, F.}, \bibinfo{author}{Peng,
  H.} \& \bibinfo{author}{Chen, Z.}
\newblock \bibinfo{title}{Novel programmable microwave photonic filter with
  arbitrary filtering shape and linear phase}.
\newblock \emph{\bibinfo{journal}{Optics express}}
  \textbf{\bibinfo{volume}{25}}, \bibinfo{pages}{9232--9243}
  (\bibinfo{year}{2017}).
\newblock \urlprefix\url{https://doi.org/10.1364/oe.25.009232}.

\bibitem{haus2012electromagnetic}
\bibinfo{author}{Haus, H.~A.}
\newblock \emph{\bibinfo{title}{Electromagnetic noise and quantum optical
  measurements}} (\bibinfo{publisher}{Springer Science \& Business Media},
  \bibinfo{year}{2012}).
\newblock \urlprefix\url{https://doi.org/10.1007/978-3-662-04190-1}.

\bibitem{liu2016frequency}
\bibinfo{author}{Liu, J.} \emph{et~al.}
\newblock \bibinfo{title}{Frequency-comb-assisted broadband precision
  spectroscopy with cascaded diode lasers}.
\newblock \emph{\bibinfo{journal}{Optics letters}}
  \textbf{\bibinfo{volume}{41}}, \bibinfo{pages}{3134--3137}
  (\bibinfo{year}{2016}).
\newblock \urlprefix\url{https://doi.org/10.1364/ol.41.003134}.

\bibitem{karpov2019dynamics}
\bibinfo{author}{Karpov, M.} \emph{et~al.}
\newblock \bibinfo{title}{Dynamics of soliton crystals in optical
  microresonators}.
\newblock \emph{\bibinfo{journal}{Nature Physics}} \bibinfo{pages}{1--7}
  (\bibinfo{year}{2019}).
\newblock \urlprefix\url{https://doi.org/10.1038/s41567-019-0635-0}.

\bibitem{leo2010temporal}
\bibinfo{author}{Leo, F.} \emph{et~al.}
\newblock \bibinfo{title}{Temporal cavity solitons in one-dimensional kerr
  media as bits in an all-optical buffer}.
\newblock \emph{\bibinfo{journal}{Nature Photonics}}
  \textbf{\bibinfo{volume}{4}}, \bibinfo{pages}{471} (\bibinfo{year}{2010}).
\newblock \urlprefix\url{https://doi.org/10.1038/nphoton.2010.120}.

\bibitem{liu2014investigation}
\bibinfo{author}{Liu, Y.} \emph{et~al.}
\newblock \bibinfo{title}{Investigation of mode coupling in normal-dispersion
  silicon nitride microresonators for kerr frequency comb generation}.
\newblock \emph{\bibinfo{journal}{optica}} \textbf{\bibinfo{volume}{1}},
  \bibinfo{pages}{137--144} (\bibinfo{year}{2014}).
\newblock \urlprefix\url{https://doi.org/10.1364/optica.1.000137}.

\bibitem{kim2019turn}
\bibinfo{author}{Kim, B.~Y.} \emph{et~al.}
\newblock \bibinfo{title}{Turn-key, high-efficiency kerr comb source}.
\newblock \emph{\bibinfo{journal}{Optics letters}}
  \textbf{\bibinfo{volume}{44}}, \bibinfo{pages}{4475--4478}
  (\bibinfo{year}{2019}).
\newblock \urlprefix\url{https://doi.org/10.1364/ol.44.004475}.

\end{thebibliography}

\end{document}


\title{Supplementary Information to: \\ Reconfigurable radiofrequency filters based on versatile soliton microcombs}

\author{Jianqi Hu}
\thanks{These authors contributed equally to the work}
\affiliation{Institute of Physics, Swiss Federal Institute of Technology Lausanne (EPFL), Photonic Systems Laboratory (PHOSL), STI-IEL, Lausanne CH-1015, Switzerland.}
 
\author{Jijun He}
\thanks{These authors contributed equally to the work}
\affiliation{Institute of Physics, Swiss Federal Institute of Technology Lausanne (EPFL), Laboratory of Photonics and Quantum Measurements (LPQM), SB-IPHYS, Lausanne CH-1015, Switzerland.}

\author{Junqiu Liu}
\affiliation{Institute of Physics, Swiss Federal Institute of Technology Lausanne (EPFL), Laboratory of Photonics and Quantum Measurements (LPQM), SB-IPHYS, Lausanne CH-1015, Switzerland.}

\author{Arslan S. Raja}
\affiliation{Institute of Physics, Swiss Federal Institute of Technology Lausanne (EPFL), Laboratory of Photonics and Quantum Measurements (LPQM), SB-IPHYS, Lausanne CH-1015, Switzerland.}

\author{Maxim Karpov}
\affiliation{Institute of Physics, Swiss Federal Institute of Technology Lausanne (EPFL), Laboratory of Photonics and Quantum Measurements (LPQM), SB-IPHYS, Lausanne CH-1015, Switzerland.}

\author{Anton Lukashchuk}
\affiliation{Institute of Physics, Swiss Federal Institute of Technology Lausanne (EPFL), Laboratory of Photonics and Quantum Measurements (LPQM), SB-IPHYS, Lausanne CH-1015, Switzerland.}

\author{Tobias J. Kippenberg}
\email[]{tobias.kippenberg@epfl.ch}
\affiliation{Institute of Physics, Swiss Federal Institute of Technology Lausanne (EPFL), Laboratory of Photonics and Quantum Measurements (LPQM), SB-IPHYS, Lausanne CH-1015, Switzerland.}

\author{Camille-Sophie Br\`es}
\email[]{camille.bres@epfl.ch}
\affiliation{Institute of Physics, Swiss Federal Institute of Technology Lausanne (EPFL), Photonic Systems Laboratory (PHOSL), STI-IEL, Lausanne CH-1015, Switzerland.}

\maketitle

\section{Derivation of RF filter responses}

The responses of comb based RF filters can be described by the discrete Fourier transform (FT) of their underlying frequency comb intensity \cite{zhu2017novel}:
\begin{equation}
H({f_{\mathrm{RF}}})\sim	\cos(2\pi^2{\phi_2}{f_{\mathrm{RF}}^2})
\sum_{k}^{} {p_{k}}\exp(j{4\pi^2}{\phi_2}k{f_m}{f_{\mathrm{RF}}})
\label{rf_filter_beta2}
\end{equation}
where $p_k$ denotes the power of each comb line, $\phi_2 = -{\beta_2}L$ is the product of the second-order dispersion $\beta_2$ of the dispersive element and its length $L$, and $f_m$ is the comb line spacing. $f_{\mathrm{FSR}} = 1/(2\pi{\phi_2}{f_m})$ is the FSR of the RF filters. The exact RF filter responses are obtained by substituting the microcomb spectral profile into Eq. \eqref{rf_filter_beta2}. For single-soliton microcomb, the optical field is given by:
\begin{equation}
 E(t) \sim  \operatorname{sech}(\frac{t}{T_0}) \otimes \sum_{n=-\infty}^{\infty}\delta(t-nT) 
\label{SS}
\end{equation}
where $ T_0 $ and $ T=1/{f_m} $ are the soliton pulse width and period, respectively. Taking the FT of Eq. \eqref{SS}, the single soliton spectrum is derived as:

\begin{equation}
\widetilde{E}(f) \sim 
\operatorname{sech}(\pi^2{T_0}f)\sum_{k=-\infty}^{\infty} \delta(f-k{f_m})
\label{SS_spectrum}
\end{equation}
This leads to the power of each comb line, or equivalently filter tap weights, being $ {p_k} \sim \operatorname{sech}^2(\pi^2k{T_0}/T)$. $ k \in Z $ is the mode index with respect to the center comb line. Disregarding  the envelope term of Eq. \eqref{rf_filter_beta2} for the moment, we can rewrite the summation part using Poisson summation formula:
\begin{equation}
\sum_{k=-\infty}^{\infty} {p_{k}}e^{j{2\pi}k\frac{f_{\mathrm{RF}}}{f_{\mathrm{FSR}}}}= f_{\mathrm{FSR}}{P(f_{\mathrm{RF}})}\otimes \sum_{n=-\infty}^{\infty}\delta(f_{\mathrm{RF}}-n{f_{\mathrm{FSR}}})
\label{poisson_summation}
\end{equation}
where $ P(f_{\mathrm{RF}}) $ is the FT of the generalized form of $ p_{k} $, at which the mode index $ k $ is substituted by an arbitrary variable $ x $, times a factor $f_{\mathrm{FSR}}$:

\begin{equation}
P(f_{\mathrm{RF}}) =  \int \operatorname{sech}^2(\frac{{\pi^2}{T_0}f_{\mathrm{FSR}}}{T}x)e^{j2\pi{f_{\mathrm{RF}}}{x}}dx   
\sim 
\frac{2\frac{T}{T_0}\frac{f_{\mathrm{RF}}}{f_{\mathrm{FSR}}}}{\operatorname{sinh}
(\frac{T}{T_0}\frac{f_{\mathrm{RF}}}{f_{\mathrm{FSR}}})
}
\equiv{G(f_{\mathrm{RF}})}
\label{linshape}
\end{equation}
where the FT of sech-squared function can be found using the residue theorem \cite{haus2012electromagnetic}, and is defined as $ G(f_{\mathrm{RF}}) $. The single-soliton based RF filter response is derived by substituting $ G(f_{\mathrm{RF}}) $ back to Eq. \eqref{poisson_summation} and incorporating the envelope term. 

Two-soliton microcomb (TSM) based RF filter responses can be obtained in a similar manner. Assume two solitons are of identical amplitude and pulse width:
\begin{equation}
E(t) \sim [\operatorname{sech}(\frac{t}{T_0}) + \operatorname{sech}(\frac{t-\frac{\alpha{T}}{2\pi}}{T_0})]\otimes \sum_{n=-\infty}^{\infty} \delta(t-nT)
\label{TSM_t}
\end{equation}
where $ \alpha $ is the azimuthal angle between two solitons, expressed in radian. The TSM spectrum can be found by FT of Eq. \eqref{TSM_t}:
\begin{equation}
\widetilde{E}(f) \sim \operatorname{sech}(\pi^2{T_0}f)(1+e^{-j\alpha{k}})\sum_{k=-\infty}^{\infty} \delta(f-k{f_m})
\label{TSM_f}
\end{equation}
We then obtain the filter tap weights as:
\begin{equation}
{p_k} \sim \operatorname{sech}^2({\pi^2}k{T_0}/T) (2+2\cos(\alpha{k}))
\label{TSM_f_power}
\end{equation}
where $k \in Z$ is the comb mode index relative to the center mode. Inserting Eq. \eqref{TSM_f_power} back to Eq. \eqref{poisson_summation} derives: 
\begin{equation}
\sum_{k=-\infty}^{\infty} {p_{k}}e^{j{2\pi}k\frac{f_{\mathrm{RF}}}{f_{\mathrm{FSR}}}} \sim  (2G(f_{\mathrm{RF}}) +G(f_{\mathrm{RF}}-\frac{\alpha}{2\pi}f_{\mathrm{FSR}}) + G(f_{\mathrm{RF}}+\frac{\alpha}{2\pi}f_{\mathrm{FSR}}))
\otimes \sum_{n=-\infty}^{\infty}\delta(f_{\mathrm{RF}}-n{f_{\mathrm{FSR}}})
\label{TSM_equation}
\end{equation}
Thus, incorporating the envelope term would eventually lead to the RF filter responses of TSM spectra.  

\section{Si$_3$N$_4$ microresonator characterization}

\begin{figure*}[t!]
    \renewcommand{\figurename}{Supplementary Figure}
    \centering{
    \includegraphics[width = 0.85\linewidth]{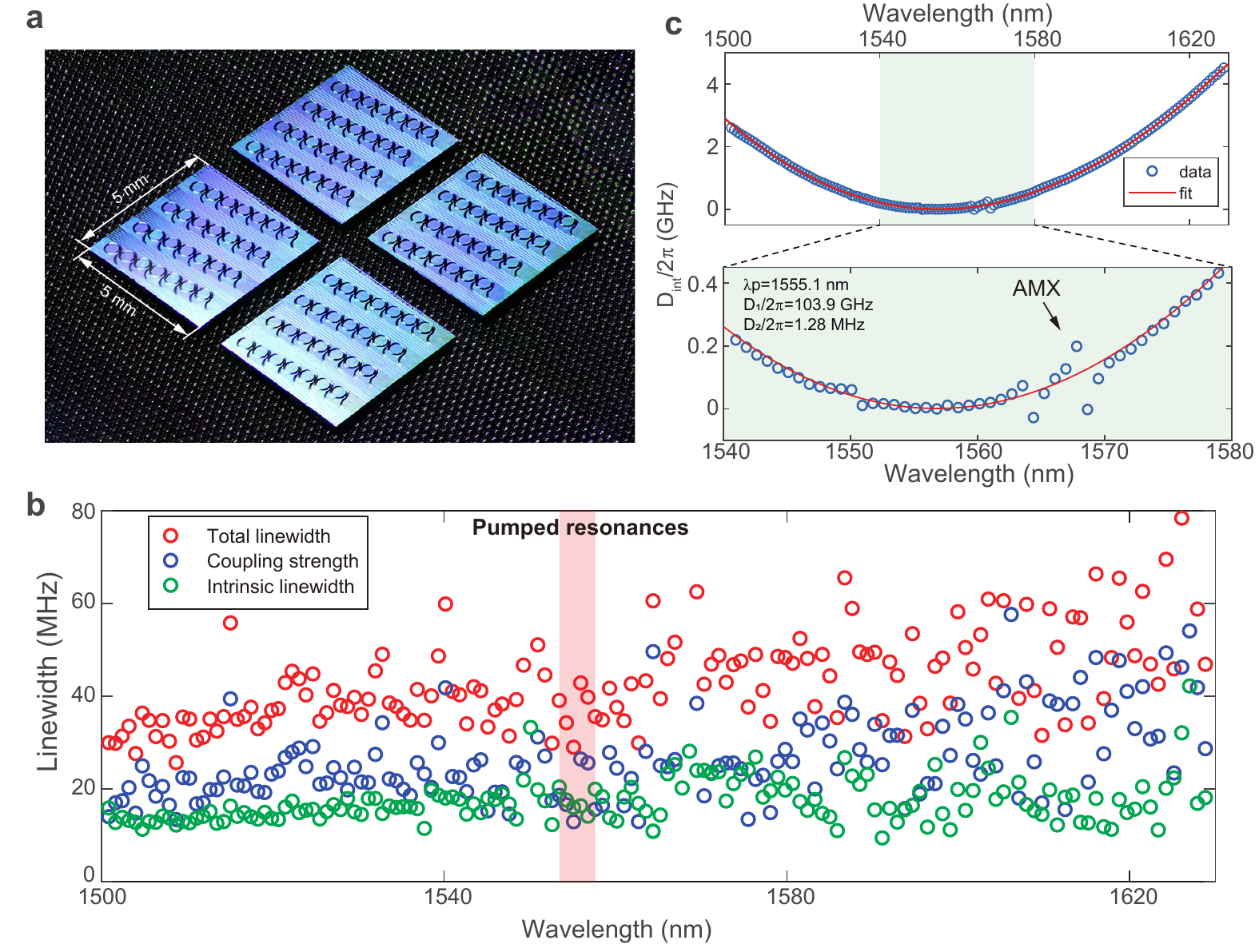}}
    \caption{\noindent \textbf{Si$_3$N$_4$ microresonator characterization.} (a) Optical image of the Si$_3$N$_4$ microresonator chips. (b) Total linewidth, coupling strength, and intrinsic linewidth of each resonance in the TE$_{00}$ mode family. The shaded area corresponds to the resonances pumped in the experiment. (c) Top: Measured integrated GVD ($D_{int}/2\pi$) of the TE$_{00}$ mode family in the microresonator, with respect to to the resonance of ${1555.1 ~{\rm nm}}$; Bottom: zoom-in of integrated GVD region between ${1540 ~{\rm nm}}$ and ${1580 ~{\rm nm}}$. Dominant AMX is observed around wavelength region of ${1565 ~{\rm nm}}$. 
    \label{chip}
}
\end{figure*}

Supplementary Figure \ref{chip}a shows a picture of Si$_3$N$_4$ microresonator chips used in the experiment. Accurate calibrated transmission spectrum of the microresonator is obtained using frequency-comb-assisted diode laser spectroscopy \cite{liu2016frequency}, covering the wavelength region from ${1500 ~{\rm nm}}$ to ${1630 ~{\rm nm}}$. Supplementary Figure \ref{chip}b illustrates the detailed properties of the resonances, i.e. intrinsic linewidths $\kappa_0/2\pi$, coupling strengths $\kappa_{\rm ex}/2\pi$, as well as the total linewidths $\kappa/2\pi = (\kappa_0+\kappa_{\rm ex})/2\pi$ of the TE$_{00}$ mode family, which are extracted from the fittings of calibrated transmission spectrum. The shaded area of Supplementary Figure \ref{chip}b denotes the experimentally accessed resonances. All these resonances show intrinsic linewidths $ \kappa_0/2\pi \approx 20 ~{\rm MHz}$, indicating the Q factors to be around $10^7$. Besides, the coupling strengths of the pumped resonances are similar to their intrinsic linewidths, implying these resonances near critical coupling condition. 

Then, the integrated GVD of the TE$_{00}$ mode family are extracted by identifying the precise frequency of each TE$_{00}$ resonance, as shown in the upper part of Supplementary Figure \ref{chip}c. It can be formulated as:
\begin{equation}
D_{\rm int}(\mu)=\omega_\mu-(\omega_0+D_1\mu)=D_2\mu^2/2+D_3\mu^3/6+...
\label{GVD}
\end{equation}
where $D_1/2\pi$ is the FSR of microresonator, and $D_n(n \in N_+ \mid n\geq 2)$ correspond to the $n$-th order dispersion coefficients. $D_{\rm int}(\mu)$ is defined as the deviation of the $\mu$-th resonance frequency $\omega_\mu/2\pi$ from the equidistant frequency grid, constructed from the FSR around the reference resonance frequency $\omega_0/2\pi$. Here, the reference resonance is chosen at $\omega_0/2\pi=192.8 ~{\rm THz}$ (i.e. $\lambda_0 = 1555.1 ~{\rm nm}$). The retrieved dispersion terms are $D_1/2\pi\approx103.9 ~{\rm GHz}$, $D_2/2\pi\approx1.28 ~{\rm MHz}$, and $D_3/2\pi\sim\mathcal{O}(1)~{\rm kHz}$. Note that operating at anomalous dispersion ($D_2>0$) is a prerequisite for soliton formation. 
The bottom part of Figure \ref{chip}c shows the zoom-in view of the integrated GVD between ${1540 ~{\rm nm}}$ and ${1580 ~{\rm nm}}$. Several resonance frequency deviations are clearly observed in the measured GVD profile, and the most dominant AMX is found around the wavelength region of ${1565 ~{\rm nm}}$. 

\section{PSC and TSM soliton steps}
\begin{figure*}[t!]
    \renewcommand{\figurename}{Supplementary Figure}
    \centering{
    \includegraphics[width = 0.75\linewidth]{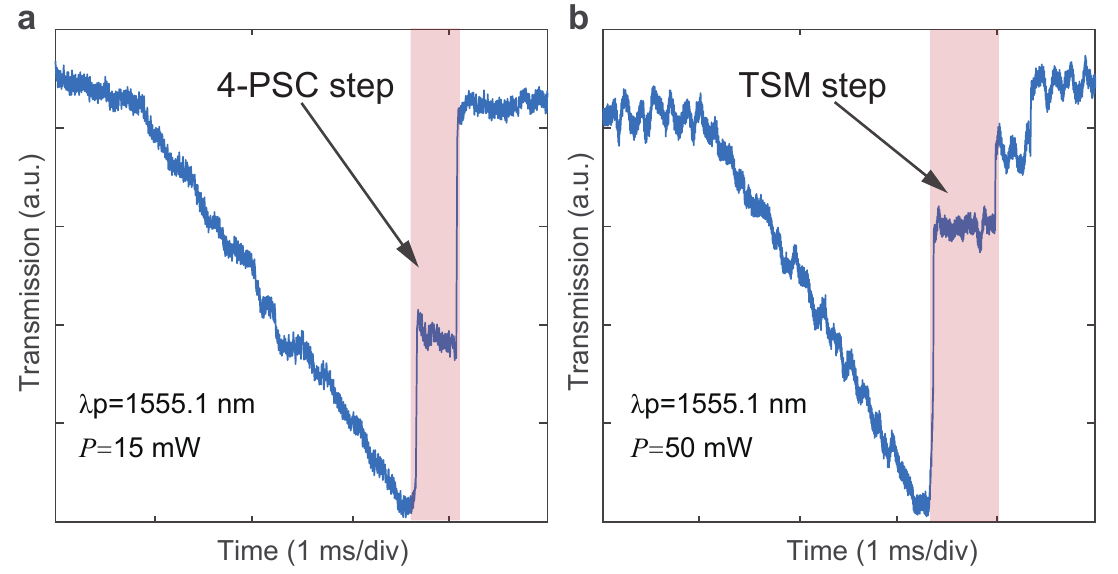}}
    \caption{\noindent \textbf{Soliton steps for PSC and TSM formations.} Transmission curves are obtained by scanning a laser across the resonance below (a) and above (b) the threshold pump power, at resonance of ${1555.1 ~{\rm nm}}$. 
    The shaded areas in (a) and (b) correspond to 4-PSC step and TSM step, respectively.
    \label{step}
}
\end{figure*}

Supplementary Figure \ref{step} depicts different soliton step formations by pumping the resonance of ${1555.1 ~{\rm nm}}$ under and above the threshold pump power level. The soliton step is manifested from the transmission of the microresonator by scanning the CW pump laser over the resonance. When the pump power is around ${15 ~{\rm mW}}$, only a single PSC step is formed, and a soliton number of $4$ can be deduced from the depth of the step. However, if the power of scanning CW laser is increased to around ${50 ~{\rm mW}}$, both two-soliton and single-soliton steps would appear. The distinct soliton steps clearly indicate two different soliton generation regimes, and are consistent with experimentally generated microcomb spectra.   

\section{TSM based RF filters of different resonances}

\begin{figure*}[t!]
    \renewcommand{\figurename}{Supplementary Figure}
    \centering{
    \includegraphics[width = 0.95\linewidth]{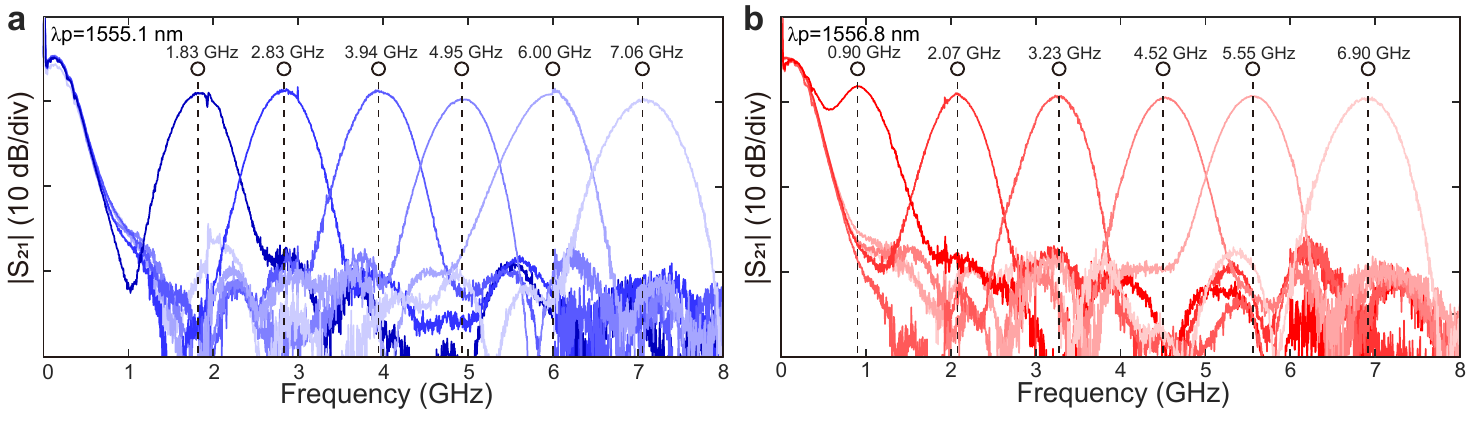}}
    \caption{\noindent \textbf{TSM based RF filter responses at resonances of 1555.1 nm and 1556.8 nm.} (a) RF filters centered at ${1.83 ~{\rm GHz}}$, ${2.83 ~{\rm GHz}}$,  ${3.94 ~{\rm GHz}}$, ${4.95 ~{\rm GHz}}$, ${6.00 ~{\rm GHz}}$, and ${7.06 ~{\rm GHz}}$ are obtained at resonance of ${1555.1 ~{\rm nm}}$. (b) RF filters centered at ${0.90 ~{\rm GHz}}$, ${2.07 ~{\rm GHz}}$,  ${3.23 ~{\rm GHz}}$, ${4.52 ~{\rm GHz}}$, ${5.55 ~{\rm GHz}}$, and ${6.90 ~{\rm GHz}}$ are obtained at resonance of ${1556.8 ~{\rm nm}}$.
    \label{response}
}
\end{figure*}

We also investigate TSM based RF filters from different pumped resonances. Supplementary Figure \ref{response}a shows the experimentally synthesized RF filters centered at ${1.83 ~{\rm GHz}}$, ${2.83 ~{\rm GHz}}$, ${3.94 ~{\rm GHz}}$, ${4.95 ~{\rm GHz}}$, ${6.00 ~{\rm GHz}}$, and ${7.06 ~{\rm GHz}}$ from the resonance of ${1555.1 ~{\rm nm}}$. While pumped at resonance of ${1556.8 ~{\rm nm}}$, the filter passband frequencies are relocated at ${0.90 ~{\rm GHz}}$, ${2.07 ~{\rm GHz}}$, ${3.23 ~{\rm GHz}}$, ${4.52 ~{\rm GHz}}$, ${5.55 ~{\rm GHz}}$, and ${6.90 ~{\rm GHz}}$ (Supplementary Figure \ref{response}b). By exploring adjacent resonances of ${1556.0 ~{\rm nm}}$, the maximum grid of TSM based RF filters is reduced to be less than ${1 ~{\rm GHz}}$. 

\begin{figure*}[t!]
    \renewcommand{\figurename}{Supplementary Figure}
    \centering{
    \includegraphics[width = 0.95\linewidth]{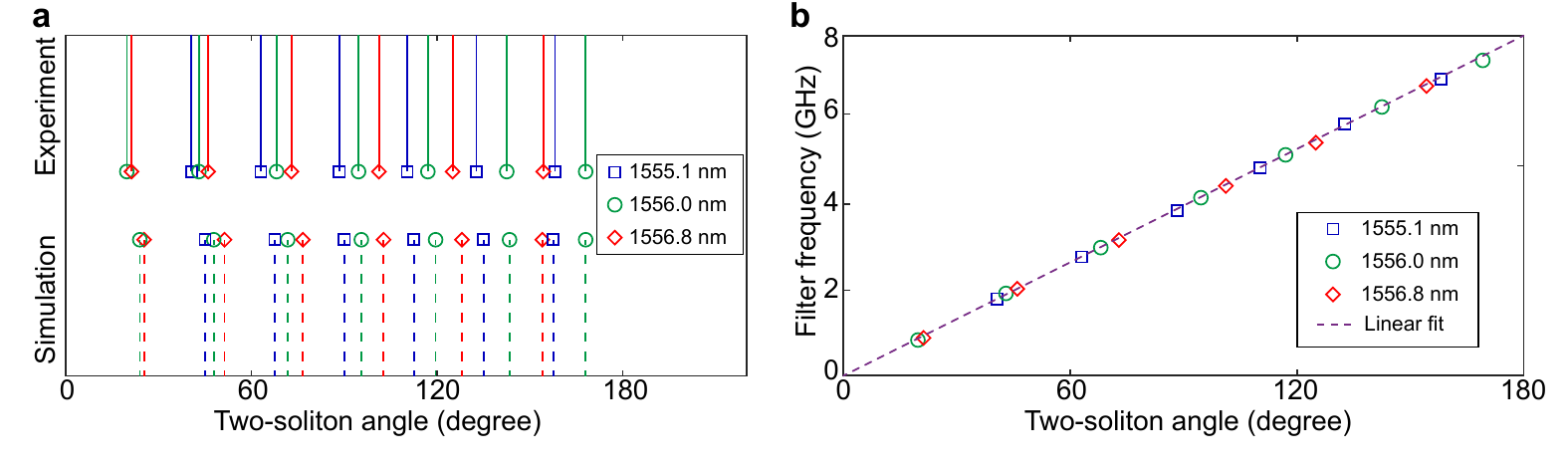}}
    \caption{\noindent \textbf{Experimental and simulated azimuthal angles of TSM spectra.} 
    (a) Experimentally retrieved (solid lines, top) and simulated (dashed lines, bottom) two-soliton angles at resonances of ${1555.1 ~{\rm nm}}$, ${1556.0 ~{\rm nm}}$, and ${1556.8 ~{\rm nm}}$. (b) The synthesized RF filter frequencies versus their underlying two-soliton angles retrieved from TSM spectra.  
    \label{angle}
}
\end{figure*}

The passband frequency shifts of RF filters arise from the variation of their underlying two-soliton azimuthal angles, which are experimentally retrieved from their TSM spectra and compared to numerical simulation (Supplementary Figure \ref{angle}a). In the simulation, the AMX position is varied from $14$-th to $16$-th away from the pump of the same strength, to resemble the change of resonances in line with experimental condition. The background modulation period is then modified according to the relative distance between the pump mode and the AMX. Besides, towards large relative soliton angle, the modification of its value through pumped resonance also becomes more prominent. This effect is simply due to the accumulated periodicity difference, and is clearly observed in both measured and simulated results. Supplementary Figure  \ref{angle}b illustrates the relation between the experimental RF passband frequencies and their corresponding TSM azimuthal angles. A good linear approximation confirms well the operation principle of proposed TSM based RF filters. 

\section{Discussion on the all-optical reconfiguration of the RF filters}
The modification of our RF filters rely fundamentally on the inherent stable and accessible DKS states. In terms of PSC microcombs, different PSC states are accessed by pumping at different resonances under the threshold power. The maximum soliton number that can be sustained in the microresonator is roughy estimated as $\sqrt{\kappa/D_2}$\cite{karpov2019dynamics}. PSC with a higher number of pulses would lead to the interaction between them. By substituting the total linewidths of the pumped resonances, the maximum PSC number in our chip is estimated around $5$. This provides a good approximation as we can achieve PSC numbers from $2$ to $4$ experimentally. We also need to point out here that not every PSC state below this predicted value is easily generated, especially for a large maximum PSC number\cite{karpov2019dynamics}. 

For the TSM spectra, the two solitons are locked to a few relative angles according to the AMX profile. Since the route to TSM undergoes chaotic region, the soliton number as well as relative distribution remain stochastic, unlike PSC formation. However, these different TSM seem to be roughly equiprobable, and all TSM states in this letter are obtained within $25$ times trial in total for each resonance. A pulse triggering technique could be envisioned in the future to deterministically switch to the TSM with targeted azimuthal angle, as has been adopted in fiber cavity \cite{leo2010temporal}. While the demand for continuous tuning of the RF passband frequency may be achieved via controlled mode interaction\cite{liu2014investigation,kim2019turn}.

\newpage
\bibliographystyle{naturemag}
\bibliography{ref}